\documentclass[useAMS,usenatbib,onecolumn]{mn2e}
\usepackage{times,graphics,epsfig}

\title{On the theory of astronomical maser. I. Statistics of maser radiation}

\author[Dinh-V-Trung]{Dinh-V-Trung\thanks{on leave from Center for Quantum Electronics, Institute of Physics and Electronics,
10 DaoTan Street, BaDinh, Hanoi, Vietnam}\\
Institute of Astronomy and Astrophysics, Academia Sinica\\ 
P.O Box 23-141, Taipei 106, Taiwan.\\
email:trung@asiaa.sinica.edu.tw
}
\begin{document}
\date{}
\maketitle
\begin{abstract}
In this paper we re-analyse the amplification process of broadband continuum radiation by 
astronomical masers in one-dimensional case. The
basic equations appropriate for the scalar maser and the random
nature of the maser radiation field are derived from basic physical principles. 
Comparision with the standard
radiation transfer equation allows us to examine the underlying assumptions
involved in the current theory of astronomical masers. Simulations are carried out 
to follow the amplification of different realisations of
the broadband background radiation by the maser.
The observable quantities such as intensity, spectral line profile 
are obtained by averaging over an ensemble of the emerging radiation
corresponding to the amplified background radiation field. Our simulations
show that the fluctuations of the radiation field inside the astronomical maser deviates 
significantly from Gaussian statistics 
even when the maser is only partially saturated. Coupling between different frequency modes and
the population pulsing are shown to have increasing importance in the transport of maser radiation 
as the maser approaches saturation. Our results suggest that the standard formulation of 
radiation transfer provides 
a satisfactory description of the intensity and the line narrowing effect in the unsaturated and
partially saturated masers within the framework of one-dimensional model. Howerver, the application of the same formulation 
to the strong saturation regime should be considered with caution.  
\end{abstract}


\section{Introduction}
Since its discovery in the interstellar medium, maser emission has been used as probe to study the
different aspects of astronomical environments. The strong maser emission has even provided detailed pictures of
the central region of distant galaxies. There had been great expectation that masers could also provide information
on the physical properties of the molecular gas at scales inaccessible by other observational means.
Theoretical studies have been carried out to find the pumping mecanisms
and the necessary conditions for the existence of masers. However, due to the strongly nonlinear nature of
masers, the relation between observed characteristics and the physical conditions of the
masing medium is difficult to determine.\\ 
At the fundamental level, the physics of astronomical maser is very similar to that of 
laboratory lasers, which are masers operating at the
optical wavelengths. Most of the properties of laboratory lasers are well described by 
the semi-classical model of Lamb (1964). The feature that distinguishes laser radiation from
classical light sources is the remarkable degree
of spatial and temporal coherence largely due to the nature of stimulated emission and the 
use of high-Q cavity and other means of stabilization. The statistics of laser radiation has been shown to 
follow Poisson-like distribution (Arechi 1965). The similarity between the laboratory laser
and the astronomical maser has prompted speculations that maser radiation might not be Gaussian.
However, previous observational attempts (Evans et al. 1972, Moran 1981) to measure the 
non-Gaussian statistics of maser radiation have proved negative.\\ 
It was recognized very early in the study of astronomical masers (Litvak 1970) that the maser radiation field is 
broadband. The coherence time set by the bandwidth of the maser 
radiation is therefore much shorter than the typical time scale of molecular response. In 
deriving the transfer equations for maser radiation, two important assumptions are used
by Litvak (1970) and later by Goldreich et al. (1973) in their study of the maser  polarization. 
The first assumption specifies that the field is
stationary and different spectral components are not correlated. The second assumption 
concerns the Gaussian distribution of the electric field components.
Goldreich et al. (1973) reason that due to the large dimension of astronomical masers, the phase
shift between waves propagating in slightly different directions is so enormous that 
the waves become completely independent and random. The electric field seen by masing 
molecules is the superposition of independent and random electromagnetic waves
and therefore will have Gaussian statistics according to the well-known central limit theorem.  
These two main assumptions allow considerable simplification in deriving the
transfer equations. 
Although several theoretical studies of maser radiation are based on these 
assumptions, their validity has not been fully examined, especially in the saturation regime
where analytical and numerical studies have been regularly carried out. 
We note here that the work by Field \& Richardson (1984) and Field \& Gray (1988) did try to
take into account the partial coherence of the maser radiation in an approximate way. Their 
analysis suggests some difference in the predicted maser intensity in comparison with the
commonly used maser theory.  
Furthermore, the recent debate on the theoretical descriptions of the polarization 
properties of astronomical masers (Watson 1994, 
Elitzur 1995) also underlines the urgent need to re-examine the foundations of the current maser theory.\\  
To simulate from first principles the amplification of radiation by a masing medium is quite
complicated. The radiation field has to be considered as an ensemble of all possible 
statistically independent realizations. Studying the amplification of maser radiation then 
requires a dramatic increase in computing power to follow the 
evolution of a large number of realizations of the incident radiation. However, from these samples we 
can study the effect of amplication on the statistical distribution of the radiation field and 
check the validity of the above-mentioned assumptions.
The physical model to describe the amplification of broadband radiation by a one-dimensional 
maser has been formulated by Menegozzi \& Lamb (1978). Using simple 
analysis they show that the Gaussian statistics can not be maintained as
the radiation is amplified and begins to affect the population inversion of the maser.
Menegozzi \& Lamb carried out the simulations in several 
cases including the medium with Dopller broadenning, which is relevent to astronomical masers. Due to 
limited computing power, they could simulate only a handful of realizations of the radiation 
field, and therefore, the results are still inconclusive. It is surprising that no further 
study has followed the work of Menegozzi \& Lamb (1978), particularly in view 
of the great computing power available today.\\
A recent work by Gray \& Bewley (2003)
also deals with the broadband nature of the maser radiation field. The
conclusion reached in their work indicates that different frequency components across the
maser line profile are phase-locked even in the partial saturation regime. 
The phase-locking (or mode-locking) effect also means that maser radiation is pulsating. 
The result is unexpected because the phase-locking of different frequency modes was not seen
in the work of Menegozzi \& Lamb (1978). Furthermore, in laboratory laser systems the mode-locking 
can only be achieved through external control either with opto-electronic components 
(Q-switch) or with passive elements such as saturable absorbers. We could not easily find a
 clear explanation on the physical mechanism giving rise to
this behaviour. We will discuss the model of Gray \& Bewley 
and compare their result with that obtained from our simulations in the following sections.\\  
In this paper we will use the formulation worked out by Menegozzi \& Lamb (1978) to study in detail
the astronomical maser and simulate from first principles the amplification of broadband background
radiation field by the maser. The characteristics of the maser
radiation will be determined directly without recourse to any approximations. 
The generalization of our model to include the vector nature of the
radiation field to study the polarization properties of masers is straightforward and will 
be described in a subsequent paper. We hope that our work will provide a better understanding 
of the underlying physical principles and the limitations of the current theory of astronomical masers.
\section{Basic theory}
\subsection{Statistical properties of radiation field}
The continuum radiation which propagates along the $z$ axis and arrives on the boundary of the masing medium is the
superposition of waves coming from many small, independent sources and 
therefore, is a random function of time. We assume that the incident radiation field $E(t)$ 
is stationary, which means that the statistical properties of the field does not change with time. We further
assume that the underlying probability distribution of random variable $E(t)$ at any instant of time
is described by a Gaussian with zero mean and variance $<|E(t)|^2>$=$\sigma^2$, where the brackets
denote the ensemble average, or the average taken over a large number of realizations of the field. 
In practice we can measure the power of the field by averaging the output of a square-law detector:
\begin{equation}
I_{\rm T} = \frac{1}{T}\int_{-T/2}^{T/2}\, |E(t)|^2\, dt
\end{equation}
Because $E(t)$ is a random process, $I_T$ defined as above for each realization of the field during
the interval $[-T/2,T/2]$ is also a random variable. As $T$ tends to infinity, $I_T$ will approach the average
power of the radiation field:
\begin{equation}
I_{\rm T} = \lim_{T\to\infty}\,\frac{1}{T}\int_{-T/2}^{T/2}\, |E(t)|^2\, dt
\end{equation}
To relate the statistical properties to time average quantities we need to restrict the random process
$E(t)$ to be ergodic, i.e ensemble average is equal to time average of any member of the ensemble. 
Justifications for this assumption constitute the famous ergodic theorem (Doob 1953). The ergodic property
is also the underlying assumption of almost all observations in radio astronomy. Therefore we can write:
\begin{equation}
<|E(t)|^2> = \lim_{T\to\infty}\,\frac{1}{T}\int_{-T/2}^{T/2}\, |E(t)|^2\, dt
\end{equation}
where the brackets denote the ensemble average. In the following sections of the paper we shall 
use the concept of spectral components of the random process 
$E(t)$. For that purpose we might formally use Fourier transform to decompose $E(t)$ 
into harmonic elements $E(\omega)$:
\begin{eqnarray}
E(t) & = & \int_{-\infty}^{+\infty}\,E(\omega)e^{-i\omega t}\, d\omega \nonumber \\
E(\omega) & = & \frac{1}{2\pi}\int_{-\infty}^{+\infty}\,E(t)e^{-i\omega t}\, dt 
\label{eq4}
\end{eqnarray} 
It becomes apparent immediately that because the random function $E(t)$ does not decay to zero at infinity and the above 
integrals do not
exist. In other words, the function $E(t)$ is not absolutely integrable. We can circumvent this difficulty by 
truncating the function $E(t)$ on a finite interval $[-T/2,T/2]$
and define a new function $E_T(t)$ which is equal to $E(t)$ over -T/2$\le\,t\,\le$T/2 and periodic outside this
interval. We can now expand $E_T(t)$ into harmonic components using Fourier series:
\begin{equation}
E_T(t) = \sum_{n=-\infty}^{n=+\infty}\, E(\omega_n) e^{-i\omega_n t}
\end{equation}
Where the angular frequency $\omega_n$=$2\pi n/T$ and 
\begin{equation}
E(\omega_n) = \frac{1}{T}\int_{-T/2}^{T/2}\, E(t)e^{i\omega_n t} dt
\end{equation}
Because $E(t)$ is a random function, the coefficients $E(\omega_n)$ are new complex random variables. 
In general, the coefficients 
$E(\omega_n)$ are independent of each other only in the limit of large T, 
i.e $<E(\omega_n) E(\omega_m)^*> = 0$ for $n\neq m$ (Root \& Pitcher 1955). For random process with 
continuum spectrum or
``white noise'', the values of $E(t)$ at different instants of time are completely uncorrelated, or 
the auto-correlation is a delta function. Therefore the coefficients $E(\omega_n)$ are Gaussian random variables and
independent even for finite values of the interval $T$ (Goldman 1953, Rice 1944, Root \& Pitcher 1955). 
Using Perceval's theorem we obtain the following relation:
\begin{equation}
\sum_{n=-\infty}^{n=+\infty}\, |E(\omega_n)|^2 = \frac{1}{T}\int_{-T/2}^{T/2} |E(t)|^2\, dt
\end{equation}  
Physically, the term $|E(\omega_n)|^2$ can be interpreted as the power of $E(t)$ contained within the 
frequency interval
$\Delta\omega$=$\omega_{n+1}-\omega_n$=$2\pi/T$ during the interval $[-T/2,T/2]$. By taking the 
ensemble average, which means averaging over a large number of realizations of the radiation field, 
we can define:
\begin{equation}
\frac{c}{8\pi}<|E(\omega_n)|^2> = S_T(\omega_n)\Delta\omega
\label{eq8}
\end{equation}
where $c$ is the speed of light. In the limit $T\to\infty$, the quantity $S_T(\omega)$ approaches
the power spectrum of the random process $E(t)$:
\begin{equation}
\lim_{T\to\infty}\, S_T(\omega) = S(\omega)
\end{equation}
The summation in the Fourier series can be formally replaced by the Fourier-Stieltjes integral in the limit that 
$T$ tends to infinity by substituting $a_n$ by $dZ(\omega)$, the increment of an orthogonal 
random process $Z(\omega)$.  
\begin{equation}
E(t) = \lim_{T\to\infty}\,\int_{-\infty}^{+\infty}\,e^{-i\omega t} dZ(\omega)
\end{equation}
where the limit should be understood in the mean square sense, namely:
\begin{equation}
\lim_{T\to\infty}\,<|E(t) - 
\int_{-\infty}^{+\infty}\,e^{-i\omega t} dZ(\omega)|^2> = 0
\end{equation}
If the random function $Z(\omega)$ is differentiable, i.e $dZ(\omega)/d\omega$ exists, the expression above 
will reduce to the normal Fourier transform. As we can see from {\bf Eq.}~\ref{eq8}, the magnitude of 
$|dZ(\omega)|$ is 
$O(\sqrt{d\omega})$, many orders of magnitude larger than $d\omega$. Hence, the derivative of $Z(\omega)$ 
does not exist. More rigorous derivation of the spectral resprentation theorem for random processes can be 
found in standard texts such as Doob (1953), Yaglom (1962) and Priestley (1981).\\
In the case of an emission line, the radiation is bandwidth limited around the frequency of the transition. 
If we take $\omega_{\rm ab}$=2$\pi\nu_{\rm ab}$, for the sake of simplicity, to be 
the central frequency of the maser line, the 
electric field of the radiation propagating along the $z$ axis can be expressed as follows:
\begin{equation}
{\rm E}(z,t) = \frac{1}{2} \left\{E(z,t)e^{-i\omega_{ab}(t - z/c)} + c.c\right\}
\end{equation}
where $c$ is the speed of light and the complex amplitudes $E(z,t)$ are slow-varying 
functions of both time and space. The complex 
function $E(z,t)e^{-i\omega_{ab}(t - z/c)}$ is called the analytic signal representation of the
electric field since the function is analytic in the upper half of the complex $\omega$ plane.
In practice we can not simulate or measure the random process $E(z,t)$ over an infinitely long interval of time, 
we shall restrict ourselves to study realizations of the radiation field during an interval $T$ 
long enough that provides adequate spectral resolution $\Delta\omega$=$2\pi/T$. 
Using the spectral representation theorem, we can expand the 
time dependent electric field in terms of Fourier series:
\begin{equation}
E(z,t)e^{-i\omega_{ab}(t-z/c)} = e^{-i\omega_{ab}(t-z/c)}\sum_{n=-\infty}^{+\infty}
E(z,\omega_n)e^{-i\omega_n(t-z/c)} 
\end{equation}
where the Fourier components are given by
\begin{equation}
E(z,\omega_n)e^{i\omega_n z/c}  = (1/T)\int_{-T/2}^{T/2} 
\left\{ E(z,t)e^{-i\omega_{ab}(t-z/c)} \right\} e^{i\omega_n t} dt
\end{equation}
with $\omega_n$= 2$\pi$n/T. For bandwidth limited signal the Fourier components $\mathbf{E}(z,\omega_n)$ 
will be zero for sufficiently large values of $n$.\\
The complex components $E(z,\omega_n)$ are random variables, whose real and imaginary parts 
($E^r$, $E^i$) are jointly independent Gaussian with zero mean and the same variance 
$<|E(z,\omega_n)|^2>$/2. The joint probability distribution is (Mandel \& Wolf 1965):
\begin{equation}
P[E(z,\omega_n)] = (1/\pi <|E(z,\omega_n)|^2>)\, {\rm exp}[-(E^r(z,\omega_n)^2 + 
E^i(z,\omega_n)^2)/<|E(z,\omega_n)|^2>]
\end{equation}
Noting that the expression for the area element 
$d^2 E$ = $d({\rm Re}\,E)d({\rm Im}\, E)$ = 
$\frac{1}{2}d|E|^2\,d\phi$ in the complex plane of $E(z,\omega_n)$
we can see that the joint density distribution for amplitude and phase of $E(z,\omega_n)$ as given
by Menegozzi \& Lamb (1978) follows directly from the above expression:
\begin{equation}
\Phi[E(z,\omega_n)] =
(1/2\pi <|E(z,\omega_n)|^2>)\, 
{\rm exp}[-|E(z,\omega_n)|^2/<|E(z,\omega_n)|^2>]
\end{equation}
with the following normalization:
\begin{equation}
\int_{0}^{2\pi}\,d\phi\, \int_{0}^{\infty}\, \Phi[E(z,\omega_n)]\,d|E(z,\omega_n)|^2 = 1
\end{equation}
The complex components $E(\omega_n)$ for each realization of the input continuum radiation field at the 
boundary of the masing medium 
can be generated with a random number generator. 
As described in Menegozzi \& Lamb (1978), if $x$ is a real random number
uniformly distributed in the interval between 0 and 1 (excluding the end points), the phase $\phi$
is simply $2\pi x$. The amplitude of the harmonic component is obtained by the relation:
\begin{equation}
|E(z,\omega_n)|^2 = -<|E(z,\omega_n)|^2>\, {\rm ln}(1\, - \, x)
\end{equation}
where $<|E(z,\omega_n)|^2>$ is the variance or the intensity of the radiation field at 
frequency $\omega_n$ and must be specified beforehand.\\      
In the studies of Goldreich et al. (1973) and Deguchi \& Watson (1990) the electric field is 
assumed to be stationary and can be 
decomposed into harmonic components using the usual Fourier tranform in similar fashion to {\bf Eq.}
~\ref{eq4}. The intensity or
Stokes parameters, if we consider the vector nature of the radiation field, are defined as the ensemble 
averaged quantities involving harmonic components at the corresponding
frequency, similar to our definition above. Although {\bf Eq.}~\ref{eq4} has only symbolic meaning, 
it can be reinterpreted using truncation technique as presented above without any change in the subsequent 
calculations.\\   
The procedure used by Elitzur (1992) is markedly different because only the decomposition of a single
realisation of the radition field into Fourier components is considered. By truncating the function 
$E(t)$ on a finite interval $[-T,T]$ and performing the usual Fourier transform, the function 
$F_{\rm T}(\omega)$ is defined as: 
\begin{equation}
F_{\rm T}(\omega)=\int_{\rm -T}^{\rm T}\, 
E(t)e^{i\omega t}\,dt
\end{equation}
The power spectrum $S(\omega)$ of the field is defined as (Elitzur 1992, {\bf Eq.} 4.1.7, 9 and 10):
\begin{equation}
\int_{0}^{+\infty}\,S(\omega)d\omega = \lim_{\rm T\to\infty}\,\frac{1}{\rm 2T}\,
\int_{\rm -T}^{\rm T}\,E(t)^2dt = \lim_{\rm T\to\infty}
\frac{\pi}{\rm T}\,\int_{-\infty}^{+\infty}|F_{\rm T}(\omega)|^2\,d\omega
\end{equation}
The last equality follows from the Perceval's theorem for the truncated function $E(t)$ over any finite
interval T. The power spectrum $S(\omega)$ of the radiation field is then identified with the limit:
\begin{equation}
S(\omega)= \lim_{\rm T\to\infty}\,\frac{2\pi}{\rm T}|F_{\rm T}(\omega)|^2 
\end{equation}
However, the existence of the average power of the stationary radiation field, i.e the limit of
$\frac{1}{\rm 2T}\,\int_{\rm -T}^{\rm T}\,E(t)^2dt$ as T tends to infinity, does not guarantee that
the limit in the above equation exists. In fact, the value of $2\pi\,|F_{\rm T}(\omega)|^2/T$, which
is called the periodogram by its originator Schuster, fluctuates wildly around the true 
power spectrum $S(\omega)$. The periodogram is usually refered as the non-consistent estimator of
the power spectrum in the signal processing literature. One way to achieve convergence is to
perform a local averaging of the periodogram (Champeney 1989):  
\begin{equation}
\lim_{\epsilon\to 0}\left\{ \lim_{\rm T\to\infty}\frac{1}{2\epsilon}\int_{-\epsilon}^{+\epsilon}
d\omega\,\frac{2\pi}{T}|F_{\rm T}(\omega)|^2 \right\} = S(\omega)
\end{equation}
This theorem shows that in order to obtain good estimate of the power spectrum, a trade-off
in spectral resolution must be made, which is of course the principle of many smoothing schemes 
to estimate the power spectrum of random processes.\\
If we want to properly define the power spectrum using time average of a single fluctuating function
$E(t)$, we need to resort to the 
techniques of generalized harmonic analysis 
(Wiener 1930). Simply speaking, to obtain the power spectrum we need to measure the auto-correlation function: 
\begin{equation}
R(\tau) = \lim_{\rm T\to\infty} \frac{1}{\rm T}\int_{\rm -T/2}^{\rm T/2}\, E(t)E^*(t\,+\,\tau)dt 
\end{equation}
The Wiener-Khintchin theorem then guarantees that the function $S(\omega)$, which can be 
identified with the power spectrum, 
exists and is related to the auto-correlation function $R(\tau)$ by the usual Fourier transform:
\begin{equation}
S(\omega) = \int_{-\infty}^{+\infty}\, R(\tau)e^{-i\omega\tau}d\tau
\end{equation}
This procedure serves as basic principle of the XF or lag correlator (Thompson et al. 2001), which is 
currently more practical than the FX design and has widespread use in radio interferometry.          
\subsection{Matter-radiation interaction in masers}
Since our aim is to simulate the spectral properties of the maser radiation, it is preferable to work from the
beginning in the frequency domain.
The basic equations are derived here for the idealized case of a scalar one-dimensional maser. That means the
masing molecules are idealized as a system with only two (upper and lower) energy levels and the properties of the maser depends
only on the $z$ coordinate, which is also the propagation axis of the maser radiation. These assumptions are 
commonly used in prevous works on the theory of astronomical masers. 
Following Menegozzi \& Lamb (1978), Goldreich et al. (1973), Deguchi \& Watson (1990) we will
derive the transfer equation for the radiation field within the framework of
the rotating-wave approximation. That means when dealing with the frequency dependence of the 
electric field and polarization
vector, we retain only the positive frequencies. The anti-resonant (negative) frequencies are ignored.
As in previous section, we express the electric field and the induced polarization vector as:
\begin{eqnarray}
{\rm E}(z,t) & = &\frac{1}{2}\{E(z,t)\, e^{-i\omega_{ab} (t\, - \, z/c)} + c.c\} \\
\lefteqn{\rm and} \hspace{5cm} \nonumber \\
{\rm P}(z,t) & = & \frac{1}{2}\{P(z,t)\, e^{-i\omega_{ab} (t\, - \, z/c)} + c.c\}
\end{eqnarray}
Where $\omega_{ab}$ is the center of the maser line, corresponding to the rest frequency of the
transition between the upper level $a$ and the ground level $b$ of the line. 
The amplitude of the field $E(z,t)$ 
and polarization $P(z,t)$ are assumed to vary slowly with time in 
comparison to the term $exp(-i\omega_{ab}\,t)$. As shown in the previous section, the amplitude of electric 
field and polarization vector 
can be expressed in terms of Fourier series for a time interval $T$, which corresponds to a spectral resolution
of $\Delta\omega\,=\,2\pi/T$:
\begin{eqnarray}
E(z,t) & = & \sum_{\rm n=-\infty}^{\rm n=+\infty}E(z,\omega_n)\, 
e^{-i\omega_n(t\, - \,z/c)} \\
\lefteqn{\rm and} \hspace{5cm} \nonumber \\
P(z,t) & = & \sum_{\rm n=-\infty}^{\rm n=+\infty}P(z,\omega_n)\, 
e^{-i\omega_n(t\, - \,z/c)}
\end{eqnarray}
The radiation transfer equation (Goldreich et al. 1973, Deguchi \& Watson 1990) can be written as follows:
\begin{equation}
\left(\frac{1}{c}\frac{\partial}{\partial\, t} \, +\, \frac{\partial}{\partial\, z}\right)\,E(z,t) =
\frac{2\pi i \omega_{ab}}{c}P(z,t)
\end{equation}
or in the frequency domain:
\begin{equation}
\frac{d}{dz}E(z,\omega_n) = \frac{2\pi i \omega_{ab}}{c}P(z,\omega_n)
\label{eq28}
\end{equation}
The polarization $P$ describes how the masing medium interacts with the radiation field 
at the frequency of the
maser line. To calculate its value we need to use the density matrix, which describes the masing medium. 
The density matrix at any position $z$ for a group of molecules moving with velocity $\upsilon$ can be written as:
\begin{equation}
\rho(z,\upsilon,t)  =  \left( \begin{array}{ll}
\rho_{\rm aa} & \rho_{\rm ab} \\
\rho_{\rm ba} & \rho_{\rm bb} \\
\end{array}
\right)
\end{equation} 
The Hermitian property of the density matrix implies that $\rho_{\rm ba} = \rho_{\rm ab}^{\rm *}$.
When there is no possible confusion we will drop the explicit dependence of the density matrix and 
the radiation field on the 
spatial coordinate $z$. To work in the frequency domain, we need to expand the density matrix elements 
into Fourier series, in a similar fashion to the electric field:
\begin{eqnarray}
\rho_{\rm ab}(z,\upsilon,t) & = & e^{-i\omega_{ab} (t\,-\, z/c)} \sum_{\rm n=-\infty}^{n=+\infty} 
[\rho_{\rm ab}(z,\upsilon,\omega_n)e^{i\omega_n z/c}] e^{-i\omega_n t} \nonumber \\
\rho_{\rm aa}(z,\upsilon,t) & = & \sum_{\rm n=-\infty}^{n=+\infty} \rho_{\rm aa}(z,\upsilon,\omega_n) 
e^{-i\omega_n (t\,-\, z/c)} \\
\rho_{\rm bb}(z,\upsilon,t) & = & \sum_{\rm n=-\infty}^{n=+\infty} \rho_{\rm bb}(z,\upsilon,\omega_n) 
e^{-i\omega_n (t\,-\, z/c)} \nonumber
\end{eqnarray}
Because the diagonal elements of the density matrix $\rho_{\rm aa}$ and $\rho_{\rm bb}$ are real, 
we have the following relations $\rho_{\rm aa}(z,\upsilon,\omega)$=$\rho_{\rm aa}(z,\upsilon,-\omega)^{\rm *}$.
The Hamiltonian of the molecules is the sum of the Hamiltonian H$_0$, representing isolated molecules, 
and the matter-radiation field interaction matrix $\mathsf{V}$. 
The well known evolution equation of the density 
matrix $\rho(z,\upsilon,t)$ (Goldreich et al. 1973, Deguchi \& Watson 1990) 
has been shown to have the following compact form by Icsevgi \& Lamb (1969) and Sargent et al. (1974):
\begin{eqnarray}
\left( \frac{\partial}{\partial t} + \upsilon \frac{\partial}{\partial z} \right) \rho_{ab}& = &
-(i\omega_{ab} + \Gamma)\rho_{\rm ab} - \frac{i}{\hbar}\mathsf{V}_{\rm ab}(\rho_{\rm bb} - \rho_{\rm
aa}) \nonumber \\
\left( \frac{\partial}{\partial t} + \upsilon \frac{\partial}{\partial z} \right) \rho_{aa}& = &
\lambda_{\rm a}(\upsilon) - \Gamma_{\rm a} \, \rho_{\rm aa} - \frac{i}{\hbar}(\mathsf{V}_{\rm ab}\rho_{\rm ba} - 
\mathsf{V}_{\rm ba}\rho_{\rm ab}) \\
 \left( \frac{\partial}{\partial t} + \upsilon \frac{\partial}{\partial z} \right) \rho_{bb}& = &
\lambda_{\rm b}(\upsilon) - \Gamma_{\rm b} \, \rho_{\rm bb} - \frac{i}{\hbar}(\mathsf{V}_{\rm ba}\rho_{\rm ab} - 
\mathsf{V}_{\rm ab}\rho_{\rm ba}) \nonumber
\end{eqnarray}
Where $\lambda_{\rm a}$ and $\lambda_{\rm b}$ are 
the pumping rates into the upper and lower levels, respectively. $\Gamma_a$ and $\Gamma_b$ are the loss rate due to 
pumping and collisional decoherence 
(Sargent et al. 1974). $\mathsf{V}_{\rm ab}$ are the matrix elements of the interaction matrix 
$\mathsf{V}$. For the sake of simplicity, we assume here that the loss rates are the same for 
the lower and upper levels. Substituting the Fourier expansion of density matrix elements into 
the above equations (except for the terms involving the interaction matrix $\mathsf{V}$ to be written 
explicitly later) and collecting term by term, we obtain:
\begin{eqnarray}
\rho_{\rm ab}(\upsilon,\omega_n) & = & -\frac{i}{\hbar}[\mathsf{V}_{\rm ab}(\rho_{\rm bb}\,-\,\rho_{\rm aa})(\upsilon,
\omega_n)]\cdot\gamma_{+}^{ab}(\omega_n,\upsilon) \nonumber \\
\label{eq32}
\rho_{\rm aa}(\upsilon,\omega_n) & = & \left\{ \lambda_{\rm a}(\upsilon)\delta_{\rm n,0} - 
\frac{i}{\hbar}[(\mathsf{V}_{\rm ab}\rho_{\rm ba} 
- \mathsf{V}_{\rm ba}\rho_{\rm ab})(\upsilon,\omega_n)]\right\}\cdot  \gamma_{+}(\omega_n,\upsilon) \\
\rho_{\rm bb}(\upsilon,\omega_n) & = & \left\{ \lambda_{\rm b}(\upsilon)\delta_{\rm n,0} - 
\frac{i}{\hbar}[(\mathsf{V}_{\rm ba}\rho_{\rm ab} 
- \mathsf{V}_{\rm ab}\rho_{\rm ba})(\upsilon,\omega_n)]\right\}\cdot  \gamma_{+}(\omega_n,\upsilon) \nonumber
\end{eqnarray}
where $\delta_{\rm n,0}$ is equal to 1 for $n$=0 and zero otherwise. The $\gamma$ functions are the Lorentzian response of the masing molecules moving at velocity 
$\upsilon$ to the radiation field and given as follows:
\begin{eqnarray}
\gamma_{\pm}^{ab}(\omega_n,\upsilon) & = & 1/\left\{\Gamma\: \pm \: i\,[\omega_{\rm ab} - (\omega_{\rm ab}\:+\:\omega_n)
\cdot(1\,-\,\frac{\upsilon}{c})] \right\} \,
\simeq 1/\left\{\Gamma\: \pm \: i\,[\omega_{\rm ab}\:\frac{\upsilon}{c}\,-\,\omega_n] \right\} 
\nonumber \\
\gamma_{\pm}(\omega_n,\upsilon) & = & 1/\left[\Gamma\: \mp \: i\,\omega_n \cdot (1\:-\:\frac{\upsilon}
{c}) \right] \, \simeq \, 1/\left[\Gamma\: \mp \: i\,\omega_n \right]
\end{eqnarray}
The functions $\gamma_{\pm}^{ab}(\omega_n,\upsilon)$ peak at 
$\omega_{\rm n}$ = $\omega_{\rm ab}\cdot\upsilon/c$, which
corresponds to the Doppler shift of the resonance frequency of the masing molecules.
We note that in arriving at {\bf Eqs.}~\ref{eq32} we have used the similar
approximation as in Section II of Menegozzi \& Lamb (1978), i.e. neglecting the term involving 
$e^{-i\omega_n (t\,-\, z/c)}\, \upsilon \frac{\partial}{\partial z}\, \rho(z,\upsilon,\omega_n)$. 
Noting further that $\upsilon\, \ll\, c$, we obtain the final expressions for the Lorentzian response
of the masing molecules. These expressions are the same as that used in Menegozzi \& Lamb (1978).  

In the frequency domaine $\omega_n$, the interaction term, i.e. the product $\mathsf{V}\cdot\rho$, 
can be written as the convolution:
\begin{equation}
\mathsf{V}\cdot\rho\,(\upsilon,\omega_n) = 
\sum_{q=-\infty}^{q=+\infty}\mathsf{V}(\omega_q)\cdot\rho(\upsilon,\omega_{\rm n-q})
\label{eq36}
\end{equation} 
The interaction matrix $\mathsf{V}$ between radiation field and the masing molecules in the dipole approximation can
be expressed as:
\begin{equation}
\mathsf{V} = -E\,\cdot\,d \simeq -\frac{1}{2} \left\{ E(z,t)e^{-i\omega_{ab}(t - z/c)} \right\} \,\cdot\,d
\end{equation}
where $d$ is the dipole moment operator for the molecule and in the last step we have made 
use of the rotating wave approximation.
For simplicity we assume that the matrix elements of the dipole moment are real, i.e 
$d_{\rm ab}=d_{\rm ab}^*=\mathbf{d}$.  
In the frequency domain the interaction term has the form:
\begin{equation}
\mathsf{V}_{\rm ab}(z,\omega_n) = -\frac{1}{2}\, \mathbf{d}\, E(z,\omega_n) 
\end{equation}
Therefore from {\bf Eqs.}~\ref{eq32} we obtain the expression for the off-diagonal element of the
density matrix:
\begin{equation}
\rho_{\rm ab}(\upsilon,\omega_n) = \frac{i\,\mathbf{d}}{2\hbar}\,
\sum_{\rm q}^{}\, E(\omega_{\rm n-q})\cdot[\rho_{\rm bb}(\upsilon,\omega_{\rm q})\:-
\:\rho_{\rm aa}(\upsilon,\omega_{\rm q})]\cdot\gamma_{+}^{\rm ab}(\omega_n,\upsilon) 
\label{eq39}
\end{equation}
The equation {\bf Eq.}~\ref{eq36} becomes:
\begin{equation}
\mathsf{V}^{\rm ab}\cdot\rho_{\rm ab}^*(\upsilon,\omega_n) = \frac{i\,\mathbf{d}^2}{4\hbar}\,
\sum_{\rm m,q}^{}\,
E(\omega_{\rm m+n})\cdot E^*(\omega_{\rm m-q})\cdot
[{\rho_{\rm bb}}^*(\upsilon,\omega_{\rm q})\: - \:{\rho_{\rm aa}}^*(\upsilon,\omega_{\rm q})]\cdot\gamma_{-}^{ab}
(\omega_{\rm m},\upsilon)
\end{equation}
We also obtain similar expression for the complex conjugate term:
\begin{equation}
{\mathsf{V}^{\rm ab}}^*\cdot\rho_{\rm ab}(\upsilon,\omega_n) = -\frac{i\,\mathbf{d}^2}{4\hbar}\,
\sum_{\rm m, q}^{}
E^*(\omega_{\rm m-n})\cdot E(\omega_{\rm m-q})\cdot
[\rho_{\rm bb}(\upsilon,\omega_{\rm q})\: - \:\rho_{\rm aa}(\upsilon,\omega_{\rm q})]\cdot\gamma_{+}^{ab}
(\omega_{\rm m},\upsilon)
\end{equation}
The population inversion of the maser can be calculated in terms of the elements of the 
density matrix as follows:
\begin{eqnarray}
\label{eq40}
\Delta\rho(\upsilon,\omega_{\rm n}) & = &\rho_{\rm aa}(\upsilon,\omega_{\rm n})\:-
\: \rho_{\rm bb}(\omega_{\rm n},\upsilon) \nonumber \\
& = & \frac{\mathbf{d}^2}{2\hbar^2} \left\{
\sum_{\rm m,\,q}E(\omega_{\rm m+n})E^*(\omega_{\rm m-q})\left[\rho_{\rm bb}^{*}(\omega_{\rm q},
\upsilon)\:-\: \rho_{\rm aa}^{*}(\omega_{\rm q},\upsilon)\right]\gamma_{-}^{ab}(\omega_{\rm m},\upsilon) 
\right. \nonumber \\
& & \left.+\:
\sum_{\rm m,\,q}E^*(\omega_{\rm m-n})E(\omega_{\rm m-q})\left[\rho_{\rm bb}(\omega_{\rm q},
\upsilon)\:-\: \rho_{\rm aa}(\omega_{\rm q},\upsilon)\right]\gamma_{+}^{ab}(\omega_{\rm m},\upsilon) 
\: + \right.\nonumber \\ 
 & & \left. (\lambda_{\rm a}(\upsilon)\:-\:\lambda_{\rm b}(\upsilon))\,\delta_{\rm n,0} 
 \vphantom{\sum_{\rm m}}\right\}\gamma_{+}(\omega_{\rm n})
\end{eqnarray}
The polarisation vector ${\rm P}(z,t)$ of the masing medium is the response of the molecules to the 
presence of the electromagnetic field. Quantum mechanically, ${\rm P}(z,t)$ is the average value of the
dipole moment operator $d$ and can be calculated as the trace of $\rho\cdot d$:
\begin{equation} 
{\rm P}(z,t) = \int\limits_{-\infty}^{+\infty}d\upsilon\, \mathbf{tr}(\rho \cdot d)
\end{equation} 
or written explicitly using the rotating wave approximation:
\begin{equation}
\frac{1}{2}P(z,t)exp\left\{-i\omega_{\rm ab}(t\, - \, z/c) \right\} 
\simeq \int_{-\infty}^{+\infty} \, d\upsilon \,\rho_{\rm ab}(z,t)\, \mathbf{d}
\end{equation}
Transforming the above equation into frequency domain and make use of {\bf Eq.}~\ref{eq39} we 
obtain:
\begin{equation}
P(\omega_{\rm n}) = \frac{i\mathbf{d}^2}{\hbar} \int_{-\infty}^{+\infty}\,d\upsilon\,
\sum_{\rm q}^{}\, E(\omega_{\rm n-q})\Delta \rho(\omega_{\rm q},\upsilon )\cdot\gamma_{+}^{\rm ab}
(\omega_{\rm n},\upsilon) 
\end{equation}
If we define the normalized profiles $\phi_\pm$, which satisfy 
$\int_{-\infty}^{+\infty}\,d\upsilon\,\phi_\pm(\omega_{\rm n},\upsilon)\, =\, 1$, as follows:
\begin{equation}
\phi_{\pm}(\omega_{\rm n},\upsilon) = \frac{\omega_{\rm ab}}{\pi {\rm c}}\cdot\gamma_{\pm}^{ab}
(\omega_{\rm n},\upsilon)
\end{equation}
the radiation transfer equation {\bf Eq.}~\ref{eq28} becomes:
\begin{equation}
\frac{d E(\omega_{\rm n})}{dz} = \frac{2\pi^2\,\mathbf{d}^2}{\hbar}\int_{-\infty}^{+\infty} d\upsilon\, 
\sum_{\rm q}\,E(\omega_{\rm n-q})\,\Delta\rho(\omega_{\rm q},\upsilon)\,\phi_{+}(\omega_{\rm n},\upsilon) 
\label{eq44}
\end{equation}
\subsection{First order approximation}
The equations derived in the previous section relating the transfer of the radiation field to 
the density matrix of masing molecules
are rather complicated. One technique to solve them is the perturbative expansion consisting in solving
the equations through a series expansion. To the first order, the populations are considered constant over
time within a given realization. Therefore the population inversion can be written simply as $\Delta\rho(\upsilon)$
with no dependence on harmonic frequency.
The density matrix element connecting the upper level to the ground 
level therefore contain electric field to the first order. The transfer equations are particularly simple.
\begin{equation}
\frac{d E(\omega_{\rm n})}{dz} = \frac{2\pi^2\,\mathbf{d}^2}{\hbar}\int d\upsilon\, 
E(\omega_{\rm n})\,\Delta\rho(\upsilon)\,\phi_{+}(\omega_{\rm n},\upsilon) 
\end{equation}
The above equation can be cast into more familiar form involving the intensity of the
radiation field per unit frequency $\nu$ ($\nu$ = $\omega/2\pi$) by defining
$I(\omega_{\rm m})\Delta\omega/2\pi\,=\,c/8\,\pi\,E(\omega_{\rm m})E^*(\omega_{\rm m})$ and
the real part $\phi^{\rm r}(\omega_{\rm m},\upsilon)$ of $\phi_{\pm}(\omega_{\rm m},\upsilon)$ :
\begin{equation}
\frac{dI(\omega_{\rm n})}{dz} = \frac{4\,\pi^2\,\mathbf{d}^2}{\hbar}\int d\upsilon\, 
I(\omega_{\rm n})\,\cdot\,\Delta\rho(\upsilon)\,\cdot\,\phi^{\rm r}(\omega_{\rm n},\upsilon)
\label{eq47}
\end{equation}
The equation for the population inversion {\bf Eq.}~\ref{eq40} in this case can be written in a particularly simple form:
\begin{equation}
\Gamma\,\Delta\rho(\upsilon) = \frac{2\mathbf{d}^2}{c\hbar^2}\sum_{\rm n}\,I(\omega_{\rm n})\,\cdot
\,\Delta\rho(\upsilon)\,\cdot \, 
\left\{\gamma_{+}^{ab}(\omega_{\rm n},\upsilon)\:+\:\gamma_{-}^{ab}(\omega_{\rm n},\upsilon)\right\}
\Delta\omega \:+\:
\Delta\lambda(\upsilon)
\end{equation}  
where $\Delta\lambda(\upsilon)$ = $\lambda_{\rm a}(\upsilon)\, -\, \lambda_{\rm b}(\upsilon)$. The 
coefficient $2\mathbf{d}^2/c\hbar^2$ can be converted to the Einstein coefficient $B$ by the relation
$8\pi^2\mathbf{d}^2/h^2c\:=\:3B/4\pi^2$.
We can simplify the equation if we define the real part of Lorentz functions $\gamma_{\pm}^{ab}(\omega_{\rm m},\upsilon)$ 
as $\gamma^{\rm r}(\omega_{\rm m},\upsilon)$:
\begin{equation}
\Gamma\,\Delta\rho(\upsilon) =  -\frac{3B}{2\pi^2}\cdot\sum_{\rm n}\,I(\omega_{\rm n})\,\cdot
\,\Delta\rho(\upsilon)\,\cdot \,\gamma^{\rm r}(\omega_{\rm n},\upsilon)\Delta\omega\:+\:\Delta\lambda(\upsilon)
\label{eq49}
\end{equation}
We note that the function $\gamma^{\rm r}(\omega_{\rm n},\upsilon)$ has the normalization 
$\int_{-\infty}^{+\infty}\,\gamma^{\rm r}(\omega_{\rm n},\upsilon)\,d\omega\,=\,\pi$. From the
radiation transfer equation we can readily derive the unsaturated optical depth of the maser as:
\begin{equation}
\tau_0(\upsilon) = \frac{4\pi^2\mathbf{d}^2}{\hbar}\,L\,\frac{\Delta\lambda(\upsilon)}{\Gamma} = 
\left(\frac{h\nu}{4\pi}\,\cdot\,3B\,\cdot\,\frac{c}{\nu}
\right)\,L\,\frac{\Delta\lambda(\upsilon)}{\Gamma}
\end{equation}
where $L$ is the length of the maser. The appearance of the factor $\nu/c$ is due to the fact that 
$\rho$($\upsilon$) has the dimension of particle number per unit 
velocity (cm$^{-3}$/cms$^{-1}$) and needs to be converted to particle number per 
unit frequency (cm$^{-3}$/Hz) with the factor $\nu/c$ in the above equation.\\
The equations {\bf Eqs.}~\ref{eq47} \& \ref{eq49} can be easily seen as identical to the equation derived by
Casperson \& Yariv (1972) to describe the amplification of optical continuum in laser amplifier with 
both homogeneous and inhomogeneous broadening. \\
The standard radiation transfer equation for the average intesity of maser emission can be recovered
by taking ensemble-averaging of the above equations and postulating that the maser medium is in steady state
with fluctuations of $\Delta\rho(\upsilon)$ independent from that of $I(\omega_{\rm m})$. Of course the last
assumption is very drastic because the population inversion is driven by the radiation field. However,
Goldreich et al. (1973) suggest that in the small signal regime the fluctuations of the population inversion
is small and has very narrow bandwidth, or the autocorrelation time much longer than that of the radiation field.
That fact can justify the assumption of ignoring the fluctuations of population inversion with respect to that of
the radiation field. Noting further that
$\gamma^{\rm r}$ and $\phi^{\rm r}$ are sharply peaked functions and satisfy the normalization 
$\sum_{\rm n}\,\gamma^{\rm r}(\omega_{\rm n},\upsilon)\Delta\omega\,=\,\pi$ and 
$\int_{-\infty}^{+\infty}\,\phi^{\rm r}(\omega_{\rm n},\upsilon)d\upsilon\,=\,1$, we
obtain the standard equation describing the ensemble anverage intensity $<I(\omega_{\rm n}>$ for maser:
\begin{equation}
\frac{d<I(\omega_{\rm n})>}{d(z/L)} = <I(\omega_{\rm n})>\,
\frac{\tau_0(\upsilon)}
{1\:+\:(3B/2 \pi \Gamma )<I(\omega_{\rm n})>} 
\label{eq50}
\end{equation}
This equation was also derived by Litvak (1970) using the integral equation approach, which makes use
explicitly of the assumption on Gaussian statistics of the radiation field.
\subsection{Change of radiation statistics}
To characterize the deviation from Gaussian statistics of the radiation field we
define the parameter $\delta$ following Menegozzi \& Lamb (1978):
\begin{equation}
\delta(\omega_{\rm m}) = \frac{<I^2(\omega_{\rm m})>\:-\:<I(\omega_{\rm m})>^2}{<I(\omega_{\rm m})>^2}
\end{equation}
For complex Gaussian random variable the value of $\delta$ is equal to unity. The evolution of $\delta$
in the unsaturated regime can be derived heuristically by taking the derivative
of $\delta$:
\begin{equation}
\frac{d\delta}{dz} = \frac{1}{<I(\omega_{\rm m})>^2}
\left(\frac{d}{dz}<I^2(\omega_{\rm m})>\:-\:
2\,\frac{<I(\omega_{\rm m})>^2}{<I(\omega_{\rm m})>}\frac{d}{dz}<I(\omega_{\rm})>\right)
\label{eq56}
\end{equation}
To estimate the change of statistics of the maser radiation field, 
we restrict ourselves to the first order approximation as shown in prevous section, mainly the equation 
{\bf Eq.}~\ref{eq47}. In this approximation the
population inversion is considered to be constant within a given realization. We note further that because 
the function $\phi^{\rm r}(\omega_{\rm n}, \upsilon)$ is a sharply peaked, the population 
inversion $\Delta\rho(\upsilon)$ at any velocity $\upsilon$ is mainly affected by
the Fourier component $\omega_{\rm n}$ of the radiation field in resonance with the molecular transition. Thus, 
we may consider the function $\phi^{\rm r}(\omega_{\rm n},\upsilon)$ as a delta function in velocity and ignore the contribution of molecules 
having velocities not in resonance with the radiation field.
In this case, the equations for the 
ensemble average $<I(\omega_{\rm m})>$ and $<I^2(\omega_{\rm m})>$ follow easily: 
\begin{eqnarray}
\frac{d}{dz}<I(\omega_{\rm m})> & = & \frac{4\pi^2\mathbf{d}^2}{\hbar}  
\,<I(\omega_{\rm m})\,\Delta\rho> \label{eq60}\\
\frac{d}{dz}<I^2(\omega_{\rm m})> & = & \frac{8\pi^2\mathbf{d}^2}{\hbar} \,<I^2(\omega_{\rm m})\,\Delta\rho>
\end{eqnarray}
It should be noted here that the fluctuations of $\Delta\rho$ are induced by the radiation field. 
In the unsaturated or small signal regime we can approximate the dependence of population inversion 
on the intensity of the radiation field as:
\begin{equation}
\Delta\rho\: \simeq \: \frac{\Delta\lambda}{\Gamma}\,
\left[1\:-\:\frac{3B}{2\pi \Gamma}\,I(\omega_{\rm m})\right]
\label{eq62}
\end{equation}
Substituting the above expressions into {\bf Eq.}~\ref{eq56} we obtain:
\begin{equation}
\frac{d\delta}{dz} = \frac{8\pi^2\mathbf{d}^2/\hbar}{<I(\omega_{\rm m})>^2}\,\frac{\Delta\lambda}{\Gamma}
\,\frac{3B}{2\pi \Gamma}\left(-<I^3(\omega_{\rm m})>\:+\:\frac{<I^2(\omega_{\rm m})>^2}{<I(\omega_{\rm m})>}\right)
\end{equation}
In the small signal regime, the initial radiation field has Gaussian statistics. As shown
in Menegozzi \& Lamb (1978), for Gaussian variables we can factorize the terms involving power 
of intensity $I$ such as $<I^3(\omega_{\rm m})>\,=\,6\,<I(\omega_{\rm m})>^3$ and 
$<I^2(\omega_{\rm m})>\,=\,2\,<I(\omega_{\rm m})>^2$. Therefore the change in statistics of the 
total radiation field seen by the molecules inside the maser is:
\begin{equation}
\frac{d\delta}{dz} = -\frac{16\pi^2\mathbf{d}^2}{\hbar} \,\frac{\Delta\lambda}{\Gamma}
\,\frac{3B}{2\pi \Gamma}<I(\omega_{\rm m})>
\end{equation}
The above equation shows that the parameter $\delta$ will decrease as the radiation is amplified by the maser. 
That means large fluctuations of the field are suppressed in comparision to smaller fluctuations.
The result has a simple physical interpretation: large fluctuations can quickly deplete the
population inversion, resulting in slower growth rate. Smaller fluctuations can grow at 
a different and faster rate, thus effectively reduce the range of intensity fluctuations.\\
We can also assess heuristically the effect of the interaction between masing molecules 
and the radiation field by comparing
the intensity predicted by the standard radiation transfer equation {\bf Eq.}~\ref{eq50} and
our model in the first order approximation described by {\bf Eqs.}~\ref{eq60}~\&~\ref{eq62}. The 
amplification of the ensemble averaged intensity of the radiation field in this approximation is:
\begin{eqnarray}
\frac{d}{dz}<I(\omega_{\rm m})> & = & \frac{4\pi^2\mathbf{d}^2}{\hbar}\,
\frac{\Delta\lambda}{\Gamma}\left(<I(\omega_{\rm m})>\,-\,
\frac{3B}{2\pi \Gamma}<I^2(\omega_{\rm m})>\right) \nonumber \\
 & = & \frac{4\pi^2\mathbf{d}^2}{\hbar}\,\frac{\Delta\lambda}{\Gamma}\left(<I(\omega_{\rm m})>\,-\,
2\: \left( \frac{3B}{2\pi \Gamma} \right)<I(\omega_{\rm m})>^2\right)
\end{eqnarray} 
The appearance of factor $2$ in the second term indicates that the fluctuations of 
radiation field can reduce the population inversion
faster then that predicted by the standard equation {\bf Eq.}~\ref{eq50}. Consequently,
our simulations will give a slightly lower ensemble average intensity under the same conditions 
for the maser in comparision to the standard maser theory. We note that previous work by
Field \& Richardson (1984) and Field \& Gray (1988), taking into account in an approximate
manner the partial coherence of the maser radiation, also reached qualitatively similar conclusion, namely
lower maser intensity appears in modest saturation regime in comparison to the standard maser theory. 
\section{Simulation of radiation transport}
In our simulation we will choose the parameters appropriate to astronomical maser.
For simplicity, we choose the loss rate $\Gamma$ = 1 s$^{-1}$ and the normalized pump rates are assumed have
a velocity dispersion $\sigma$ with the difference in pump rates $\Delta\lambda(\upsilon)\,=\,
\exp(-\upsilon^2/\sigma^2)$. In our simulation we use 200 modes with a frequency resolution of 
$\Delta\omega=0.75\,\Gamma$ or 0.75 s$^{-1}$ around the maser line and we assume that this frequency
range covers the range $-\sigma$ to $+\sigma$ in the velocity domain. The actual velocity resolution $\Delta\upsilon$ 
will depend on the frequency of the maser line, as $\Delta\upsilon$ = $c/\omega_{\rm ab}\Delta\omega$.
For the maser line such as the 1612 MHz OH maser, $\Delta\upsilon$ is approximately 2 cm s$^{-1}$. The
corresponding value for the velocity dispersion is 200 cm s$^{-1}$
Although the velocity dispersion of the maser line in our simulations is much smaller than in real 
astronomical masers, the number of frequency modes and the bandwidth are large enough, i.e. the bandwidth
of 150 s$^{-1}$ is much greater than the loss rate of 1 s$^{-1}$,
to capture the main features of the broadband radiation field produced by the astronomical masers.   
Since we deals with only partially saturated maser, we consider
only 15 harmonic components of the population inversion $\rho(\omega_n)$. As shown later, 
the number of harmonic components
is enough to capture the pulsation of the molecular population inversion.
We generate the background continuum radiation in a simimlar way as Menegozzi \& Lamb (1978) 
using random generator ran2 from Press et al. (1992). The phase of
electric field components is random and uniformly distributed over the interval 0 to 2$\pi$.
For the sake of simplicity, we assume that 
the amplitude of different modes has a Gaussian distribution with zero mean and
variance $<I(\omega_{\rm n})\Delta\omega>\,=\,$1 on scale of $2h\nu^3/c^2$. On this scale the
intensity has the unit of the photon occupation number.
The intensity of the maser shown in all figures is also of the form $I(\omega_{\rm n})\Delta\omega$, where
$I(\omega_{\rm n})$ is the intensity per unit frequency $\nu$ as defined in previous section. 
We use fourth-order Runge-Kutta method with a fixed 
step $h\,=\,$0.01 to integrate the {\bf Eq.}~\ref{eq44}.
Each realization is evolved through the maser using the rate equations {\bf Eqs.}~\ref{eq40} together with the
radiation transfer equation {\bf Eq.}~\ref{eq44} 
and we record the emergent radiation for later analysis.\\
We carry out our simulations in the partially saturated regime, which is likely relevant to most astronomical
masers. The choice is also necessary because we consider only a limited number of harmonic components of the
molecular population inversion. In the saturated regime the fluctuation will be stronger and thus require 
the consideration of a larger number of harmonic components. The amplification of the maser is specified 
by the unsaturated optical depth $\tau_0(\upsilon\,=\,0)$ at the line center. A value of 
$\tau_0(\upsilon\,=\,0)$ = 22 is used throughout in our
simulations. The spontaneous transition rate between the upper and lower maser levels is 
taken to be 10$^{-9}$ s$^{-1}$.
The Einstein coefficient $B$ is related to the spontaneous transition rate by the 
well-known relation $A\,=\,2h\nu^3/c^2\,B$.
\section{Discussion}
The new ingredient of our model is the explicit treatment of the radom radiation field and the 
inclusion of the molecular population pulsation and the coupling between
different modes of the radiation field. We can examine each of these effects on the propagation
of electromagnetic waves in the maser. 

\begin{figure}
\includegraphics[width=16cm]{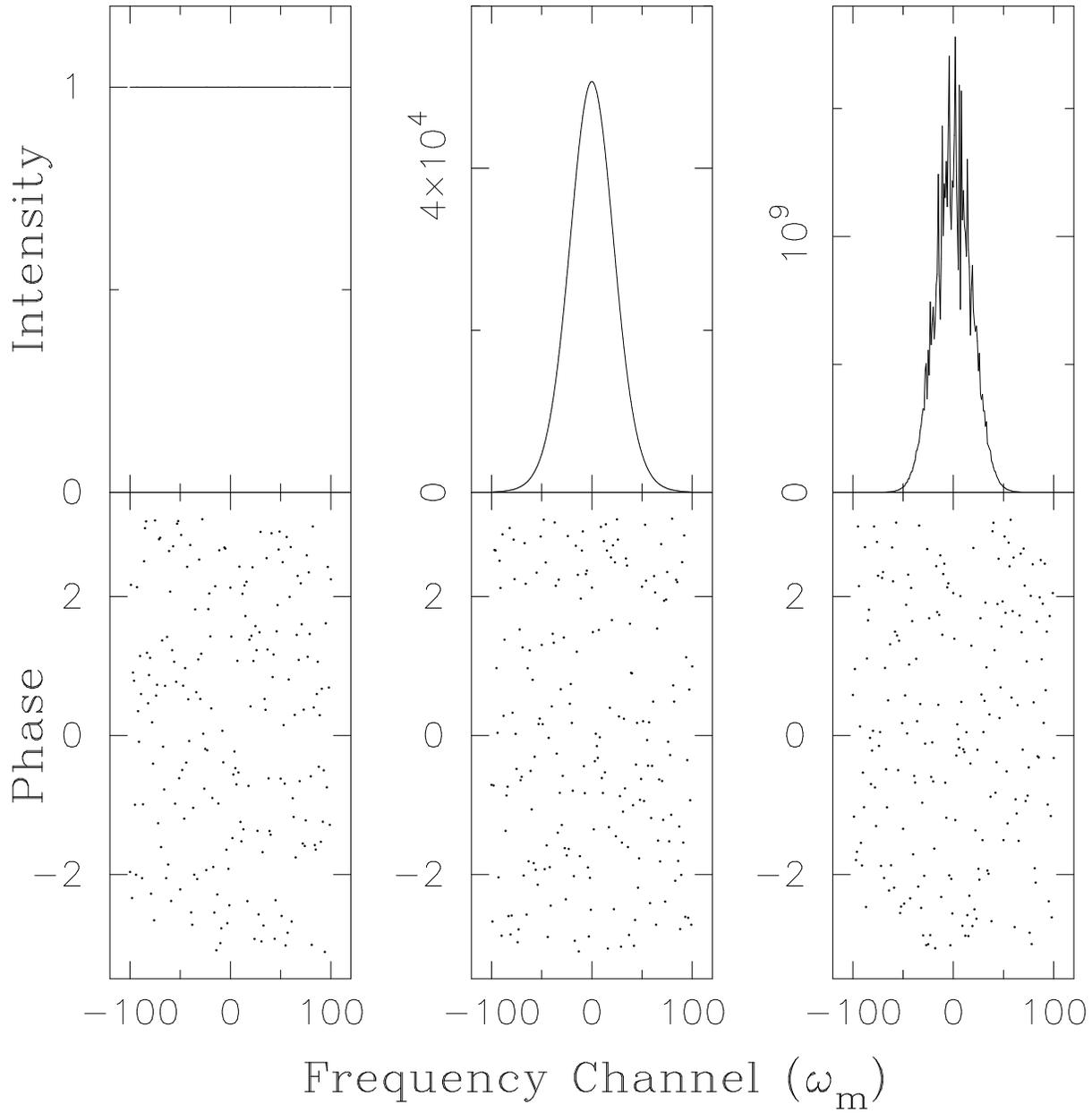}
\caption{Evolution of a realization of the radiation field with different modes 
having the same amplitude but random phases. Intensity is on the scale of 2$h\nu^3/c^2$.
{\bf Left frame} shows the initial intensity and 
phases of the radiation field. {\bf Middle frame} shows
the intensity and phases of the field half-way into the maser medium, corresponding to
$\tau_0(\upsilon=0)$ = 11. {\bf Right frame} 
shows the intensity and phases of radiation field
at the output of the maser.}
\label{fig1}
\end{figure}

\begin{figure}
\includegraphics[width=16cm]{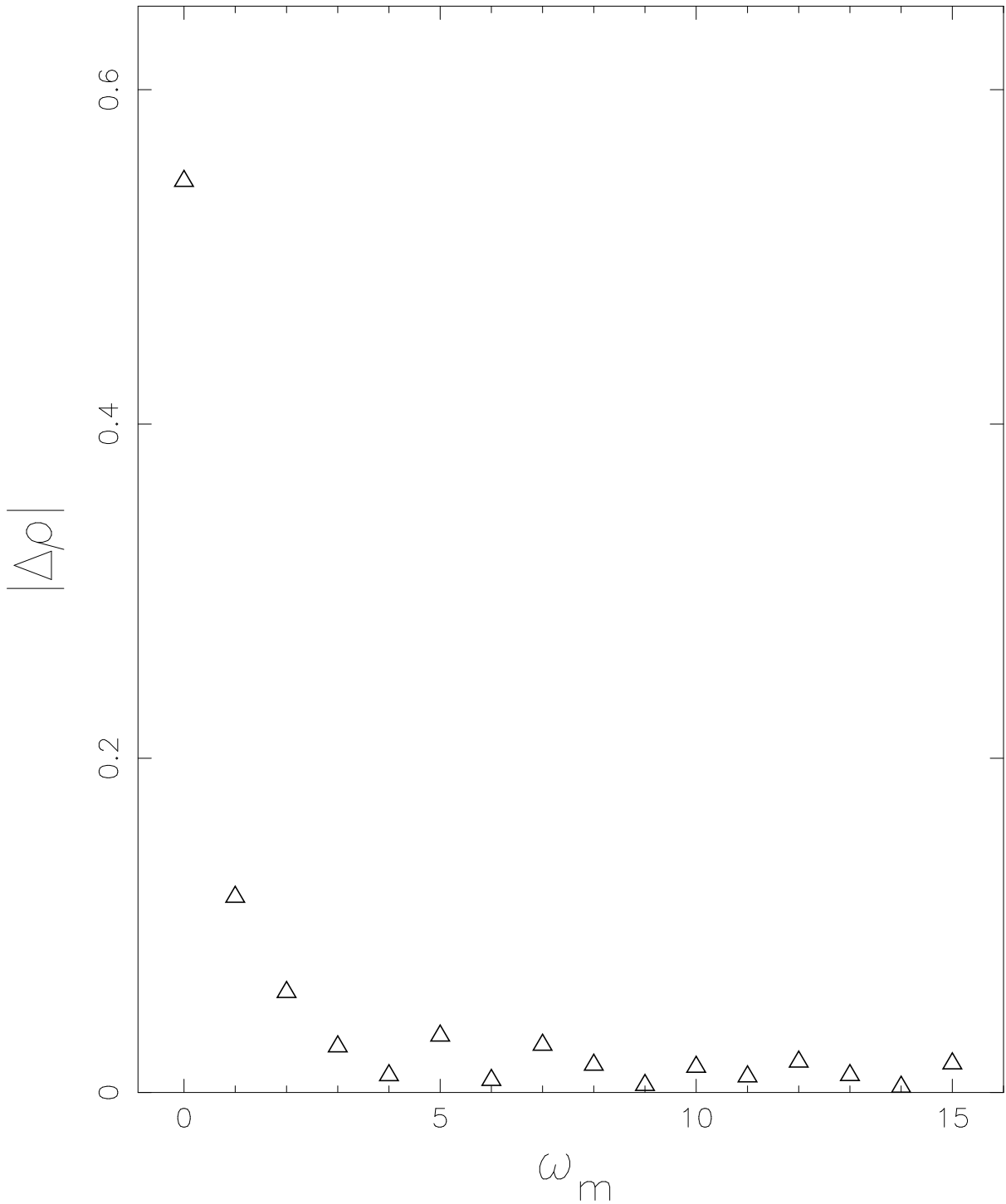}
\caption{Amplitude of Fourier components of the population pulsation with velocity 
$\upsilon$\,=\,0 at the output of the maser medium ($z$ = $L$), corresponding
to the simulation presented in Fig.~\ref{fig1}.}
\label{fig2}
\end{figure}

\begin{figure}
\includegraphics[width=16cm]{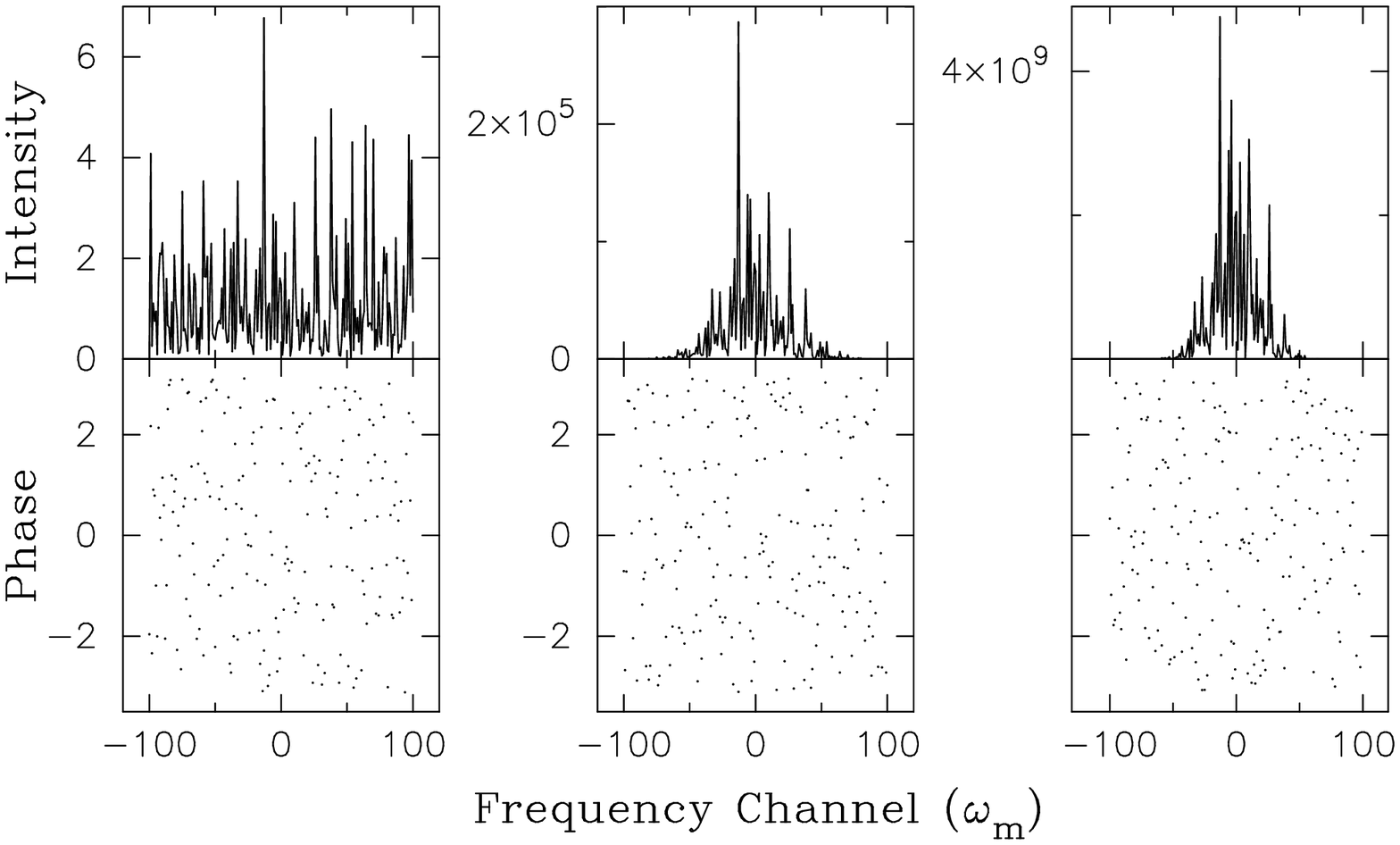}
\caption{Evolution of a realization of the radiation field with different modes 
having the random amplitude and random phases. Intensity is on the scale of 2$h\nu^3/c^2$.
{\bf Left frame} shows the initial intensity and 
phases of the radiation field. {\bf Middle frame} shows
the intensity and phases of the field half-way into the maser medium, corresponding to
$\tau_0(\upsilon=0)$ = 11. {\bf Right frame} 
shows the intensity and phases of the emerging radiation field
at the output of the maser.}
\label{fig3}
\end{figure}

If we neglect the population pulsation, as can be seen from {\bf Eq.}~\ref{eq44} only the intensity
of the waves enters into the radiation transfer equation. Thus different realizations with initial constant amplitude
but random phases are indistinguishable and we expect them to evolve similarly. However, when population pulsation is
taken into account, different modes can interact and the random phases can induce fluctuations in the amplitude
of the waves.
In {\bf Fig.}~\ref{fig1} we show the evolution 
of a particular realization of the field whose modes have constant amplitude but random phases. Initially the mode-coupling 
is negligible and the intensity
across the maser line half way into the maser medium, which corresponds to the optical depth of $\tau$/2, varies 
smoothly according to the change 
of the optical depth with frequency. The usual line-narrowing effect can be clearly seen because the initial FWHM of the maser line due to 
inhomogeneous (Doppler) broadening is 2$\sigma$ln2$\,\sim\,$140 frequency channels and the FWHM of the laser line
at the mid-point of the laser (see the middle frame of {\bf Fig.}~\ref{fig1}) is only around 50 frequency channels. 
As the intensity grows different
modes of the radiation field inside the homogeneous Lorentzian profile can couple and drive the molecular population
pulsations. The emergent maser radiation fluctuates widely around the line center where this effect is strongest 
({\bf Fig.}~\ref{fig1}). The maser in this case is only partially saturated with the saturation parameter R/$\Gamma\,\sim\,0.8$, where
$R$ is the simulated emission rate.
In {\bf Fig.}~\ref{fig2} we show the amplitude of the Fourier components for the population pulsations at the line center. 
The dominant component is at zero frequency or the average of population inversion. The amplitude of other Fourier 
components remains small and decreases at higher pulsation frequency. Nevertherless the population pulsation and mode-coupling 
are effective in inducing the fluctuations of the 
amplified maser radiation. We can also see from {\bf Fig.}~\ref{fig1} that the phases of different modes remain 
random as the waves propagate through the maser and experience amplification. No mode-locking or phase correlation 
is seen in our simulations. We also checked the simulation results shown in Menegozzi \& Lamb (1978) and could not
find any evidence for phase correlation between different modes of the radiation field.
Thus, our results do not corroborate the suggestion of Gray \& Bewley (2003) that there is a
strong phase correlation between frequency modes or mode-locking even for partially saturated maser. The likely reason 
for the difference is that the formulation of Gray \& Bewley (2003), which follows closely 
that of Yariv (1989) for the amplification of monochromatic (single mode) laser emission, 
instead of the broadband random radiation field. 
The steady state constraint imposed on {\bf Eq.}~28 of their paper leads to the presence in
subsequent equations of time-dependent terms, 
which are supposed to be eliminated by the same procedure. Of course, the problem can be avoided if 
their equation {\bf Eq.}~28 is expanded into Fourier series as we have done here.\\
In {\bf Fig.}~\ref{fig3} we show the input radiation field with both random amplitude and phase of different modes 
are random. We also show in {\bf Fig.}~\ref{fig3} the radiation field at the mid-point of the maser and the emergent radiation field.
As can be seen in the figure, only modes around the line center are amplified strongly. Because of the
nonlinear amplification, the relative intensity fluctuations between different mode changes as the radiation is amplified
in the maser medium. That is the case for the relatively strong mode seen close to the line center in the initial input field
(see {\bf Fig.}~\ref{fig3}). When the intensity of the radiation field is still small, i.e. the maser is unsaturated,
all the modes are amplified by the same factor as evident in the middle frame of {\bf Fig.}~\ref{fig3}. However, as the
maser approaches saturation, the adjacent modes, which are initially weaker, grow faster. As a result, the relative
intensity fluctuation is reduced in the output radiation field. We also note that the phases between
modes are still completely random as seen in {\bf Fig.}~\ref{fig3}.

\begin{figure}
\includegraphics[width=16cm]{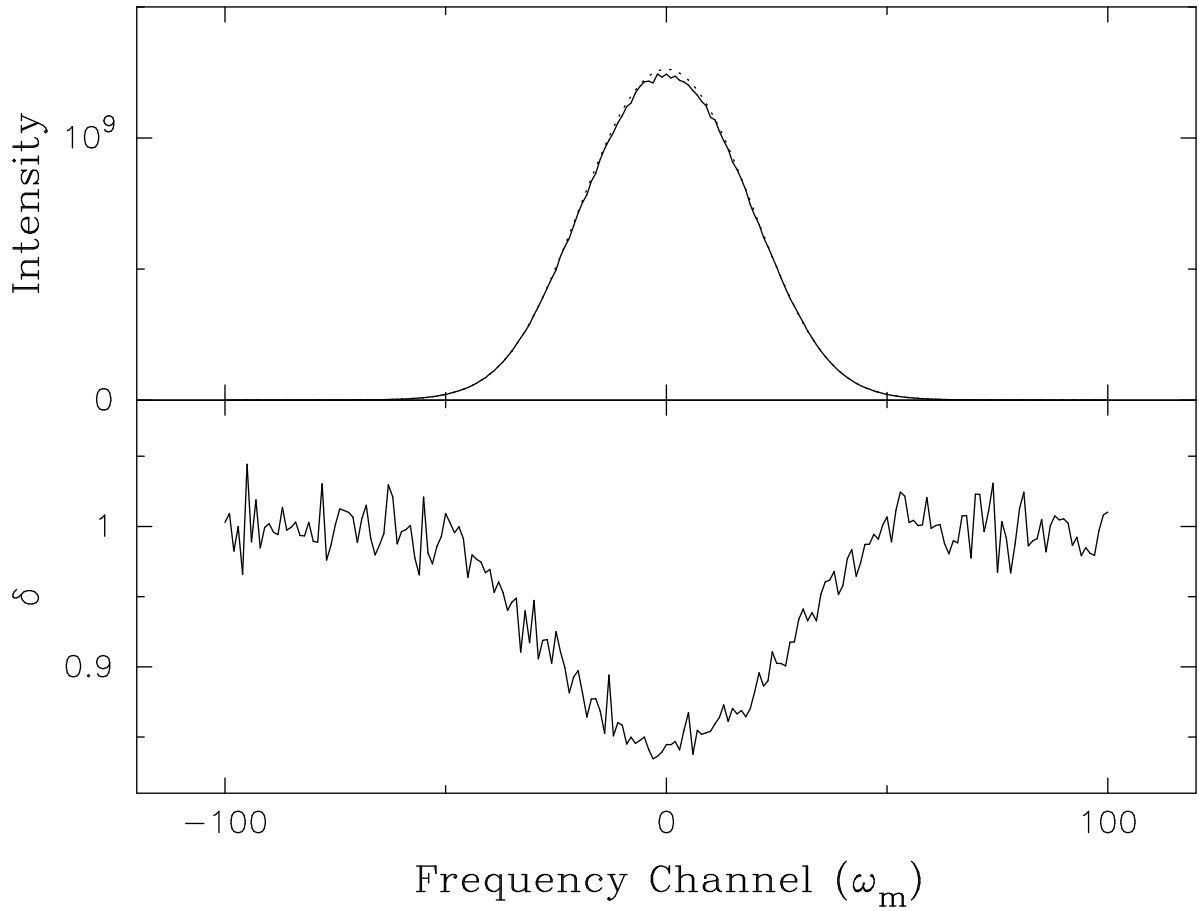}
\caption{{\bf Upper frame:} ensemble average power spectrum taken over 18000 independent 
samples of the emerging radiation field. The maser emission calculated using the standard 
radiation transfer equation is shown in dotted line. Intensity is on the scale of 2$h\nu^3/c^2$.
{\bf Lower frame:} ensemble average of parameter $\delta$, i.e. the departure from Gaussian statistics, 
taken over the same number of independent samples of the emerging radiation field.
\label{fig4}}
\end{figure}

We have followed the amplification of an ensemble of 18000 realizations of the incident radiation field through the
maser. By taking the ensemble average of the output field we could build up the observable intensity profile of the
maser line and the statistics of the radiation field.      
The parameter $\delta$, which measures the deviation from Gaussian statistics, is shown in {\bf Fig.}~\ref{fig4}.
For the first time we are able to determine the statistics of the maser radiation directly from numerical simulations. 
Even in our case of partial saturation
with R/$\Gamma\:\sim\:0.8$ at the line center, the statistical distribution of the total radiation field 
deviates quite significantly from Gaussian. The decrease of $\delta$ at the line center suggests that 
intensity fluctuations diminish with saturation and the statistical distribution of the radiation 
field becomes more centrally peaked. As discussed previously, the non-Gaussian statistics is the direct consequence of the
non-linear interaction between masing molecules and the radiation field. We could expect that the deviation 
will be even more pronounced as the maser becomes fully saturated.\\  
We also compare the ensemble averaged intensity of the total electric field taken over 18000 realizations of the field
(see {\bf Fig.}~\ref{fig4}) with the prediction of the standard radiation transfer equations {\bf Eq.}~\ref{eq50}. The two 
calculations match closely with each other in the line wings. At the center of the maser line our 
simulation produces a slightly lower ensemble average intensity 
than that given by the standard theory, consistent with our heuristic analysis in previous section.
Therefore in the unsaturated and partially saturated regime the standard theory of astronomical maser can produce results in 
good agreement with that from our more realistic treatment.
However, in saturated masers where population 
pulsation and mode-coupling are expected to be much stronger than considered here and the radiation field deviates even more from 
Gaussian statistics, the usual approach used by Litvak (1970) and Goldreich et al. (1973) are no longer applicable.
As a result, previous 
calculations using the standard radiation transfer equations and carried out in the full saturation regime 
with R/$\Gamma\,\gg\,10$ (Western \& Watson 1984, Deguchi \& Watson 1990, Nedoluha \& Watson 1990) might not give 
accurate description of the maser radiation and should be considered with cautions. We also note that fully saturated maser
is of purely theoretical interest because real astronomical masers are unlikely to be found in this regime due to various 
constraints including the pumping of masers. Although it is difficult to determine the degree of saturation for
astronomical masers, the modelling results of Watson et al. (2002) suggest that recent observations of water maser 
features with Gaussian spectral line profile (Sarma et al. 2001) can only be explained if the maser is unsaturated
with R/$\Gamma\,<\,1/3$.\\
The deviation from Gaussian statistics seen in our simulations can be measured in practice 
if we can observe an isolated maser source with frequency resolution comparable to the
homogeneous line width $\sim\Gamma$. When the frequency resolution is much larger than the homogenous bandwidth, the measured
electric field is the sum over many independently random harmonic components and the resulting statistics is expected 
to become Gaussian. Large bandwidth is used in previous attempts (Evans et al. 1972, Moran 1981) 
to measure the statistics of the maser radiation and that might contribute to the negative results, although other complications 
such as the presence of several independent maser soures inside the radio telescope beam might preclude any such detection.\\  

Finally we should emphasize that our formulation of the maser theory, which is based on the work of Menegozzi \& Lamb (178), 
as well as the standard theory of maser of Litvak (1970) and Goldreich et al. (1973) are all based on the 
one-dimensional idealization of the maser medium. Thus
the maser radiation field consists of plane waves propagating along the same direction. In real astronomical masers, the
masing medium is a three-dimensional object. The radiation field at any point in the medium is the superposition of waves
propagating in many different directions. The formaluation of the maser theory in three dimensions,
although similar in principle, will be more complicated. The abovementioned results and conclusions from our simulations 
will need to be reconfirmed in the future with fully three dimensional calculations. In addition, the spontaneous emission
by the masing molecules has been omitted in our model. This simplification is quite common in the theoretical studies 
of astronomical masers. An explicit treatment of the spontaneous emission will require the reformulation of the maser theory
using quantum electrodynamics, probably along the line presented in the classic work of Sargent et al. (1974) and
clearly is out of the scope of our paper. However, we note that, in a number of cases the astronomical masers are known 
observationally to amplified the background continuum source (such as the central red giant star or the
AGN). Thus it's reasonable to expect that the omission of the spontaneous emission will not change qualitatively the results of
our model. 
\section{Conclusion}
The explicit incorporation of the broadband random radiation field and the molecular population pulsation into our 
treatment of astronomical maser has proved very important to investigate the properties of maser emission. 
We have shown clearly the effect of mode coupling in driving the molecular population pulsation. The 
amplification of background radiation by maser is accompanied by the change of statistics of the radiation field 
in which the large fluctuations are suppressed relative to smaller fluctuations. Our simulation results suggest
that the standard radiation transfer equation provides a good description of the maser properties such as intensity
and line-narrowing in the unsaturated and partially saturated regime. Howerver, the application of the same equation to study 
masers in the strong saturation regime should be considered with caution. Further studies on the effect of mode coupling 
and population pulsation in saturated masers are needed and we hope to address them in a future publication. 

\section*{Acknowledgments}

We would like to thank the referee, Dr. M.D. Gray, for insightful and constructive comments that help to improve
greatly the presentation of our paper. This research has made use of 
NASA's Astrophysics Data System Bibliographic Services
and the SIMBAD database, operated at CDS, Strasbourg, France.

\newpage

\begin{thebibliography}{}
%
\bibitem[Arechi (1965)]{Arechi} Arechi, F.T., 1965, Phys. Rev. Letters, 15, 912 
\bibitem[Born \& Wolf (1999)]{Born} Born, M., Wolf, E., Principles of optics, 7th ed., 1999, 
Cambridge University Press 
\bibitem[Casperson \& Yariv (1972)]{Cas72} Casperson, L.W., Yariv, A., 1972, IEEE J. Quantum Electronics
vol. QE-8, 80
\bibitem[Champeney (1989)]{Cha89} Champeney D. C., 1989, A handbook of Fourier theorems, Cambridge University Press
\bibitem[Deguchi \& Watson (1990)]{De90} Deguchi, S., Watson, W.D., 1990, ApJ 354, 649
\bibitem[Doob (1953)]{Doob} Doob, J. L., Stochastic processes, 1953, Wiley, New York
\bibitem[Elitzur (1992)]{El92} Elitzur, M., Astronomical maser, 1992, Kluwer Academic Publishers
\bibitem[Elitzur (1995)]{el95} Elitzur, M., 1995, ApJ 440, 345
\bibitem[Evans et al. (1972)]{Ev72} Evans N.J., Hills R.E., Rydbeck, O.E.H., Kollberg E., 1972, Phys. Rev A,
6, 1643
\bibitem[Field \& Richardson (1984)]{field84} Field D., Richardson I.M., 1984, MNRAS 211, 799
\bibitem[Field \& Gray (1988)]{field88} Field D., Gray M.D., 1988, MNRAS 234, 353
\bibitem[Goldreich et al. (1973)]{Go73} Goldreich P., Keeley, D.A., Kwan, J.Y., 1973, ApJ 
179, 111 
\bibitem[Goldman (1953)]{Gold} Goldman, S., Information theory, 1953, Prentice-Hall, New York
\bibitem[Gray \& Bewley (2003)]{Gray} Gray, M.D., Bewley, S.L., 2003, MNRAS 344, 493
\bibitem[Icsevgi \& Lamb (1969)]{Ic69} Icsevgi, A., Lamb W.E.Jr, 1969, 
Phys. Rev, 185, 517
\bibitem[Lamb (164)]{lamb64} Lamb, W.E.Jr., 1964, Phys. Rev 134, A1429
\bibitem[Litvak (1970)]{Litvak} Litvak, M.M., 1970, Phys. Rev A, 2, 2107
\bibitem[Mandel \& Wolf (1965)]{Mandel} Mandel, L., Wolf, E., 1965, Rev. of Mod. Phys., 37, 231
\bibitem[Menegozzi \& Lamb (1978)]{Me78} Menegozzi, L.N., Lamb, W.E.Jr, 1978, Phys. Rev A,
17, 701
\bibitem[Moran (1981)]{Moran} Moran, J., 1981, BAAS, 15, 508
\bibitem[Nedoluha \& Watson 1990]{ne90} Nedoluha, G.E., Watson, W.D., 1990, ApJ 354, 660
\bibitem[Press et al. (1992)]{Press} Press, W. H., Teukolsky, S. A., Vetterling, W. T., Flannery, B. P., 
Numerical recipes in fortran 77, 1992, Cambridge University Press, New York
\bibitem[Priestley (1981)]{priestley} Priestley, M. B., Spectral analysis and time series, 1981, Academic Press  
\bibitem[Rice (1945)]{rice} Rice, S. O., Bell Syst. Tech. J., 1944, 23, 282; 1945, 24, 46; reproduced in Noise
and Stochastic Processes, N. Wax, Ed., Dover, New York, 1954
\bibitem[Root \& Pitcher (1955)]{Root} Root, W.L., Pitcher, T.S., 1955, Ann. Math. Stat., 26, 313
\bibitem[Sargent et al. (1974)]{Sa74} Sargent, M., Scully M., Lamb, W.E.Jr, 1974, Laser
Physics, Addison-Wesley, Reading, Massachusette
\bibitem[Sarma et al. (2001)]{Sa01} Sarma, A.P., Troland, T.H., Romney, J.D., 2001, ApJ, 554, L217
\bibitem[Thompson et al. (2001)]{Thompson} Thompson, A. R., Moran, J. M., Swenson, G. W. Jr., 2001, 
Interferometry and synthesis in radio astronomy, John Wiley \& Sons, New York
\bibitem[Watson (1994)]{W} Watson, W.D., 1994, ApJ 424, L37
\bibitem[Watson et al. (2002)]{W02} Watson, W.D., Sarma, A.P., Singleton, M.S., 2002, ApJ 570, L37 
\bibitem[Western \& Watson (1984)]{We84} Western, L.R., Watson, W.D., 1984, ApJ 285, 158
\bibitem[Wiener (1930)]{Wiener} Wiener, N., 1930, Acta. Math. 55, 117 
\bibitem[Yaglom (1962)]{Ya} Yaglom, A. M., An introduction to the theory of stationary random functions, 
1962, Prentice-Hall, New York
\bibitem[Yariv (1989)]{Yv} Yariv, A., Quantum electronics third edition, 1989, Wiley 
\end{thebibliography}
\end{document}